\documentclass[10pt,A4paper,conference]{IEEEtran}
\usepackage{mathpple}
\usepackage{times}
\usepackage{amsmath}  % Define \boldsymbol (in amsbsy too) and align
\usepackage{amssymb}  % Define \mathbb (in amsfonts too)
\usepackage{mathrsfs} % Define \mathscr, a script font
\usepackage{cite}  
\usepackage{upref}
\usepackage{amsfonts}
\usepackage{latexsym}
\usepackage{graphicx}
\usepackage{verbatim}

\usepackage{stmaryrd}
\usepackage{amscd}

\newcommand{\zero}{{\mathbf 0}}

\newcommand{\floorenv}[1]{\left\lfloor #1 \right\rfloor}

\newcommand{\sbinom}[2]{\left[ \begin{array}{c} #1 \\ #2 \end{array} \right] }
\newcommand{\sbinomq}[2]{\sbinom{#1}{#2}_q }

\newcommand{\cA}{{\cal A}}
\newcommand{\cB}{{\cal B}}
\newcommand{\cC}{{\cal C}}

\newcommand{\cG}{{\cal G}}

\newcommand{\cI}{{\cal I}}

\newcommand{\cN}{{\cal N}}

\newcommand{\cP}{{\cal P}}

\newcommand{\cS}{{\cal S}}

\newcommand{\cT}{{\cal T}}

\newcommand{\highsup}[1]{\raisebox{0.35ex}{\kern 1pt $\scriptstyle {#1} $}}

\makeatletter
\DeclareRobustCommand{\sbinom}{\genfrac[]\z@{}}
\makeatother
\newcommand{\G}[2]{\sbinom{{#1}\kern-1pt}{{#2}\kern-1pt}}
\newcommand{\Gq}[2]{\sbinom{{#1}\kern-0.25pt}{{#2}\kern-0.25pt}}

\newcommand{\sbinomtwo}[2]{\sbinom{#1}{#2}_2 }
\newcommand{\B}{{\mathbb B}}
\newcommand{\F}{{\mathbb F}}

\newcommand{\Fqn}{\smash{{\mathbb F}_{\!q}^n}}

\newcommand{\GF}{{\rm GF}}

\newcommand{\C}{{\mathbb C}}
\newcommand{\Z}{{\mathbb Z}}

\newcommand{\al}{\alpha}

\newcommand{\sP}{\cP}
\newcommand{\sG}{\cG}

\newcommand{\Gr}{\smash{{\sG\kern-1.5pt}_q\kern-0.5pt(n,k)}}
\newcommand{\Grtwo}{\smash{{\sG\kern-1.5pt}_2\kern-0.5pt(n,k)}}
\newcommand{\Gkone}{\smash{{\sG\kern-1.5pt}_q\kern-0.5pt(n,k_1)}}
\newcommand{\Gktwo}{\smash{{\sG\kern-1.5pt}_q\kern-0.5pt(n,k_2)}}
\newcommand{\Ps}{\smash{{\sP\kern-2.0pt}_q\kern-0.5pt(n)}}

\newcommand{\deff}{\mbox{$\stackrel{\rm def}{=}$}}

\newtheorem{theorem}{Theorem}

\newtheorem{lemma}{Lemma}

\newtheorem{cor}{Corollary}

\newtheorem{example}{Example}

\begin{document}
\vspace{-1.5cm}

\title{\vspace{-0cm}A New Construction for Constant Weight Codes\vspace{-0.3cm}}

\author{

\authorblockN{Tuvi Etzion}
\authorblockA{Department of Computer Science\\
Technion - Israel Institute of Technology\\
Haifa 32000, Israel,
email: etzion@cs.technion.ac.il} \and
\authorblockN{Alexander Vardy}
\authorblockA{Dept. of Electrical and Computer Engineering\\
University of California San Diego\\
La Jolla, CA 92093, USA,
email: avardy@ucsd.edu}\vspace{-3cm}}

\maketitle
\begin{abstract}
A new construction for constant weight
codes is presented. The codes are constructed from $k$-dimensional
subspaces of the vector space $\Fqn$. These subspaces form a
constant dimension code in the Grassmannian space
$\cG_q(n,k)$. Some of the constructed
codes are optimal constant weight codes with parameters not known
before. An efficient algorithm for error-correction is given for
the constructed codes. If the constant dimension code has an efficient
encoding and decoding algorithms then also the constructed constant
weight code has an efficient encoding and decoding algorithms.
\end{abstract}

\vspace{0cm}

\vspace{-0.3ex}
%@@@@@@@@@@@@@@@@@@@@@@@@@@@@@@@@@@@@@@@@@@@@@@@@@@@@@@@@@@@@@@@@@@@@@@@@%
%                                                                        %
%         1. INTRODUCTION                                                %
%                                                                        %
%@@@@@@@@@@@@@@@@@@@@@@@@@@@@@@@@@@@@@@@@@@@@@@@@@@@@@@@@@@@@@@@@@@@@@@@@%
\section{Introduction}
\vspace{-0.4ex}
\label{sec1}

\noindent\looseness=-1 \PARstart{C}{onstant} weight codes were
extensively studied. These codes have various important
applications, e.g.~\cite{CHT,CSW,Imm91,AGM}.
Let an $(n,d,w)$ code be a binary \emph{constant weight code} of
length $n$, constant weight $w$ for the codewords, and minimum
Hamming distance~$d$. Let
$A(n,d,w)$ be the maximum number of codewords in an $(n,d,w)$
code. The
quantity $A(n,d,w)$ was also a subject for dozens of papers,
e.g.~\cite{Jon62,BSSS,AVZ}. Some optimal
constant weight codes can be translated to other combinatorial
structures such as Steiner systems, difference families, and
Hadamard matrices and these were also investigated in the context
of their coding theory applications~\cite{MWS,Rot96} and
combinatorial
designs~\cite{Wil72}.

Some exact values of the quantity $A(n,d,w)$, like
those derived from Steiner systems, are known. But, usually
the exact value is not known. There are also some efficient constant weight
codes~\cite{Knu86}, and also a general efficient encoding
algorithm for some classes of codes~\cite{Cov73}. There are also
some error-correction for other classes~\cite{CHT},
but these are exceptional and usually given either to relatively
small codes or codes which are not interesting from minimum
distance point of view. The goal of this paper is to present a new
construction for constant weight codes. Our
construction produces for some $n$, $d$, $w$, some new $(n,d,w)$ codes which are
larger than other known $(n,d,w)$ codes with the same parameters. We
design efficient encoding/decoding algorithms and
also efficient error-correction algorithm for our codes.

The paper is organized as follows. In
Section~\ref{sec:construction} we present the construction of our
codes. In Section~\ref{sec:analysis} we analysis the codes
obtained from our construction. We present three examples of known
optimal codes which are also derived by our construction. We
present new optimal constant weight codes not known before which
are generated by our construction. In Section~\ref{sec:encode}
we discuss encoding/decoding and
error-correction algorithms for our codes. Conclusion is given in
Section~\ref{sec:last}.

\vspace{0.00ex}
%@@@@@@@@@@@@@@@@@@@@@@@@@@@@@@@@@@@@@@@@@@@@@@@@@@@@@@@@@@@@@@@@@@@@@@@@%
%                                                                        %
%   2. Bounds on the Size of Codes in Projective Space                   %
%                                                                        %
%@@@@@@@@@@@@@@@@@@@@@@@@@@@@@@@@@@@@@@@@@@@@@@@@@@@@@@@@@@@@@@@@@@@@@@@@%
\section{Construction for Constant Weight Codes}
\vspace{.5ex} \label{sec:construction}

In this section we present the new construction for constant weight
codes. Constructions
in~\cite{OMK,MOKL} and in~\cite{EtVa11a} are special cases of our construction.
The main
ingredients for our construction are constant dimension codes.
These codes got lot of interest recently due to their application
in error-correction for network coding~\cite{KK08}. Many papers
have been considered this topic recently,
e.g.~\cite{SKK,KoKu08,EtSi09,EtVa11,BEOVW,TMBR}.
Given a nonnegative integer $k \le n$, the set of all subspaces of
$\F_q^n$ with dimension $k$ is known as a \emph{Grassmannian},
and usually denoted by $\cG_q(n,k)$. It turns out
that~the natural~measure of distance in $\cG_q(n,k)$ is given by~\cite{KK08,SKK,EtVa11}
\vspace{-0.08in}
\begin{equation}
\label{eq:d-def} d(U,\!V) \,\ \deff\ \dim U + \dim V -2 \dim\bigl(
U\, {\cap} V\bigr)
\vspace{-0.07in}
\end{equation}
for all $U,V \,{\in}\, \cG_q(n,k)$. We say that $\C
\kern1pt{\subseteq}\kern1pt \cG_q(n,k)$ is an $[n,d,k]_q$
\emph{code in the Grassmannian} if $d(U,\!V) \ge d$ for
all $U,V$ in $\C$.
Let
$\cA_q(n,d,k)$ be the maximum number of codewords in an $[n,d,k]_q$
code. The input for our construction is a
constant dimension code $\C$. The cosets of each
subspace from $\C$ are transferred into words with the same length and
weight. In other word, this is a construction which transfers {\it
from dimension to weight} and hence we will call it Construction
FDTW.

One representation of a $k$-dimensional subspace $X$ of $\F_q^n$
(or any of its $q^{n-k}$ cosets in $\F_q^n$, including $X$) is by the $q^k$
vectors of length $n$ which are contained in $X$ (or its coset, respectively).
Let $\F_{q^n}$ be a finite field with $q^n$ elements, where~$q$ is
a power of a prime number,
and let $\alpha$ be a primitive element in~$\F_{q^n}$.
It is well-known that there is an isomorphism between~$\F_{q^n}$ and
$\F_q^n$, where the \emph{zero} elements are mapped into each other,
and $\alpha^i \in \F_{q^n}$, $0 \leq i \leq q^n-2$, is mapped into its
$q$-ary $n\text{-}$tuple representation in $\F_q^n$, and vice versa.
Using this mapping, a $k$-dimensional subspace $X$
of~$\F_q^n$ is represented by the corresponding $q^k$ elements of~$\F_{q^n}$ in $X$.
Throughout this paper we will not distinguish in all
places between the two representations
and the vector representation will coincide
in many places with the finite field representation.

Similarly to the two possible representations of codewords
in a constant dimension code there are two possible
representations for codewords in an $(n,d,w)$ code. The first representation
in as a binary word (vector) of length $n$. The second
representation is as a $w$-subset of the $n$-set $\{ 1,2, \ldots ,n \}$,
where a codewords contains the $w$ nonzero entries in the codeword.

We will also need the definition of a
characteristic vector $ch(A)$ for a subset $A=\{ a_1 , a_2 ,
\ldots ,a_m \}$ of $\Fqn$. The characteristic vector function
induces a mapping from the set of all $m$-subsets of $\Fqn$ into
the set of all binary vectors of length $q^n$ and weight $m$,
where $ch(A) = (c_0,c_1,\ldots,c_{q^n-1})$ is given by
\vspace{-0.00in}
$$
c_i = 1 ~\text{if}~ \al^i {\in}\kern1pt A ~~ \text{and}~~ c_i = 0
~\text{if} ~\al^i \notin A,~~ 0 \leq i \leq q^n-2 ,
$$
$$
c_{q^n-1} = 1 ~\text{if}~ 0 {\in}\kern1pt A ~~~ \text{and}~~
c_{q^n-1} = 0 ~\text{if} ~0 \notin A~.
$$
Let $X$ be a subset of $\F_q^n$ and $\beta \in \Fqn$. The
addition $\beta + X$ is defined as the addition of $\beta$ to each
element of $X$. If $X = \{ \gamma_1 , \gamma_2 , \ldots , \gamma_m
\}$ then $\beta + X \deff \{ \beta + \gamma_1 , \beta + \gamma_2 ,
\ldots , \beta +\gamma_m \}$. Note that $\beta$ and each $\gamma_i$,
$1 \leq i \leq m$, is a vector of length $n$ over $\F_q$
(or equivalently an element in $\F_q^n$).

\vspace{0.1cm}

\noindent {\bf Construction FDTW:}

\vspace{0.1cm}

Let $\C$ be an $[n,d,k]_q$ code. Given a codeword $X=\{ 0 ,
\alpha_1 , \ldots , \alpha_{q^k-1} \} \in \C$ we form a set of
codewords $\cC_X$ as follows:
$$
\cC_X \deff \{ ch(\{ \beta , \beta + \alpha_1 , \beta + \alpha_2 ,
\ldots , \beta + \alpha_{2^k-1} \}) ~:~ \beta \in \F_q^n \}~.
$$
The codewords of $\cC_X$ are the
cosets of the the $k$-dimensional subspace $X$. Therefore,
$|\cC_X|= q^{n-k}$. We define a constant weight code $\cC$ as
union of these characteristic vectors obtained from
all the codewords of $\C$, i.e.,
\begin{align*}
\cC \deff \bigcup_{X \in \C} \cC_X = \{ ch(\{ \beta , \beta + \alpha_1
, \beta + \alpha_2 , \ldots , \beta + \alpha_{2^k-1} \}) ~: \\ \{
0 , \alpha_1 , \ldots , \alpha_{q^k-1} \} \in \C , ~ \beta \in
\F_q^n \}~.
\end{align*}
\begin{theorem}
\label{thm:binary} If $\C$ is an $[n,d=2t,k]_q$ constant dimension
code then the code $\cC$ obtained by Construction FDTW is a $(q^n ,
2 \cdot q^k - 2 \cdot q^{k-t} ,q^k)$
code with $q^{n-k} | \C |$ codewords.
\end{theorem}
\begin{proof}
The length of the code $\cC$ and the weight of its codewords are
obvious. Since the number of cosets of a~$k\text{-}$dimensional~subspace
in $\Fqn$ is $q^{n-k}$ it follows that the number of codewords in
$\cC$ is $q^{n-k} | \C |$. Assume that the
minimum distance of $\cC$ is less than $2 \cdot q^k - 2 \cdot
q^{k-t}$. Then there exist two distinct codewords in $\cC$ which
have at least $q^{k-t}+1$ entries with {\it ones} located on the
same position numbers in both codewords. Hence, the intersection
of the corresponding $q^k$-subsets $X,~Y$ of $\Fqn$ has at least
$q^{k-t}+1$ elements. Clearly $X$ and $Y$ are not cosets of the
same codeword of $\C$ since all the distinct cosets of the same
codeword are disjoint. Let $-\beta \in X \cap Y$, $C(X) \deff \beta +
X$, and $C(Y) \deff \beta + Y$. Since ${\bf 0} \in C(X) \cap C(Y)$,
where ${\bf 0}$ is the allzero vector, it follows that $C(X),~C(Y)
\in \C$. $|X \cap Y| \geq q^{k-t}+1$ implies that $|C(X) \cap C(Y)
\geq 2^{k-t}+1$. Therefore, $X$ and $Y$ (and hence $C(X)$ and
$C(Y)$) share at least $k-t+1$ linearly independent elements,
i.e., $\dim (C(X) \cap C(Y)) \geq k-t+1$ and hence by~(\ref{eq:d-def})
we have $d(C(X),C(Y)) \leq k + k -2(k-t+1)=2t-2$ which
contradicts the minimum distance of $\C$.

Thus, the minimum distance of $\cC$ is $2 \cdot q^k - 2 \cdot
q^{k-t}$.
\end{proof}

For a binary code $\cC$
(in the Hamming space) of length $n$ (not necessarily
constant weight) the \emph{shortened} code by the
coordinate $i$, $\cC_b^i$, $b \in \F_2$, is defined by
\begin{align*}
\cC_b^i = \{ (c_0,\ldots,c_{i-1},c_{i+1},\ldots,c_{n-1} ) ~: \\
(c_0,\ldots,c_{i-1},b,c_{i+1},\ldots,c_{n-1} ) \in \cC   \}~.
\end{align*}
Hence, for each $b$, $b \in \F_2$, we can form $n$ shortened
codes. It is readily verified that the length of each shortened
code is~$n-1$ and its minimum distance is the same as the minimum
distance of $\cC$. The size of the shortened code might depend on
the coordinate of the shortening. Since the cosets of a subspace
over $\F_q^n$ form a partition of $\F_q^n$ it follows that the
size of the shortened codes from Construction FDTW does not depend
on the coordinate of the shortening. The size of the code is different if
$b$ is \emph{zero} or \emph{one}. By applying the shortening operation
on the codes obtained by Construction FDTW we can easily infer
the following theorem.

\begin{theorem}
\label{thm:shortq} If $\C$ is an $[n,d=2t,k]_q$ constant dimension
code then there exist a $(q^n -1 , 2 \cdot q^k -
2 \cdot q^{k-t} ,q^k-1)$ constant weight code of size $|\C|$ and a $(q^n -1
, 2 \cdot q^k - 2 \cdot q^{k-t} ,q^k)$ constant
weight code of size $(q^{n-k} -1) | \C |$.
\end{theorem}

A construction of some specific
$(q^n -1  , 2 \cdot q^k - 2
\cdot q^{k-1} ,q^k)$ codes of size $(q^{n-k} -1) |\C|$
was given in~\cite{OMK} and of
some specific $(q^n -1 , 2 \cdot q^k - 2 ,q^k)$ codes
of size $(q^{n-k} -1) |\C|$
was given in~\cite{MOKL}. Their
constructed codes were introduced as
optical orthogonal codes. In the following
section we will explain when the code obtained by Construction
FDTW will be an optical orthogonal code.

\vspace{0.2ex}
%@@@@@@@@@@@@@@@@@@@@@@@@@@@@@@@@@@@@@@@@@@@@@@@@@@@@@@@@@@@@@@@@@@@@@@@@%
%                                                                        %
%   3. Constructions of Codes in Projective Space                        %
%                                                                        %
%@@@@@@@@@@@@@@@@@@@@@@@@@@@@@@@@@@@@@@@@@@@@@@@@@@@@@@@@@@@@@@@@@@@@@@@@%
\section{Analysis on the Size of the Codes}
\label{sec:analysis}

In this section we examine some codes obtained by
Construction FDTW.
For this we need the $q$-ary {\it Gaussian
coefficient} $\sbinomq{n}{\ell}$ defined as follows
(see~\cite[p. 325]{vLWi}):
\begin{align*}
\sbinomq{n}{\ell} = \frac{(q^n-1)(q^{n-1}-1) \cdots
(q^{n-\ell+1}-1)}{(q^\ell-1)(q^{\ell-1}-1) \cdots (q-1)} ~, ~~
\sbinomq{n}{0} =1 ~.
\end{align*}
Another two concepts which will appear in our discussion
are Steiner systems and $q$-analog of Steiner system.
A \emph{Steiner system} $S(t,w,n)$ is a collection $\cB$ of
$w$-subsets taken from an $n$-set $\cN$ such that
each $t$-subset of $\cN$ is contained in exactly on element
of $\cB$.
A Steiner system $S(t,w,n)$ is also an
$(n,d,w)$ code of size \smash{$M = \binom{n}{t}/\binom{w}{t}$} and
$d = 2(w-t+1)$. A \emph{$q$-analog Steiner system}
$\cS_q[t,k,n]$ is a collection $\B$ of
$k$-dimensional subspaces taken from $\Fqn$ such that each
$t$-dimensional subspace of $\Fqn$ is contained in exactly one
element of $\B$. It can be easily verified that a $q$-analog Steiner
system $\cS_q [t,k,n]$ is an $[n,d,k]_q$ code
of size \smash{$M = \sbinomq{n}{t}/\sbinomq{k}{t}$} and $d =
2(k-t+1)$. $q$-analog Steiner system $\cS_q[1,k,n]$ exists if and only if
$k$ divides $n$. They are also known as spreads in projective
geometry~\cite[p. 330]{vLWi}.
%Such spreads were studied in many papers, e.g.~\cite{Bu,HoPa}.

Let $n=s k$, $r = \frac{q^n-1}{q^k-1}$, and let $\alpha$ be a
primitive element in GF($q^n$). For each $i$, $0 \leq i \leq r-1$,
we define
$$H_i = \{ \alpha^i , \alpha^{r+i} , \alpha^{2r+i}
, \ldots , \alpha^{(q^k-2)r+i} \}.$$ The set $\{ H_i ~:~ 0 \leq i
\leq r-1 \}$ is a $q$-analog Steiner system $\cS_q[1,k,n]$, i.e., an $[n,
2k,k]$ code of size $\frac{q^n-1}{q^k-1}$.

Only recently the first known $q$-analog Steiner system $\cS_q[t,k,n]$,
with $1 < t < k < n$ was constructed~\cite{BEOVW}. This is a
$q$-analog Steiner system $\cS_2[2,3,13]$. Construction FDTW was
applied on this system (as was described in~\cite{EtVa11a}) to obtain
a Steiner system $S(3,8,8192)$.

\begin{example}
\label{ex:Steiner} Let $\C$ be an $[n, 2,2]_2$
code of size $\sbinomtwo{n}{2}$ which consists of all 2-dimensional
subspaces from $\F_2^n$. By Construction FDTW we form a
$(2^n, 4,4)$ code~$\cC$ of size $2^{n-2} \sbinomtwo{n}{2}$. $C$ consists
of the codewords of weight four in the extended Hamming code of length
$2^n$~\cite{MWS}, i.e., a Steiner system $S(3,4,2^n)$.
\end{example}

\begin{example}
\label{ex:Hadamard} Let $\C$ be an $[n,
2,n-1]_2$ code which consists of all the $(n-1)$-dimensional subspaces
from $\F_2^n$. Applying Construction FDTW on $\C$ we form a
$(2^n , 2^{n-1} , 2^{n-1})$ code~$\cC$ of size $2^{n+1}-2$.
If we join to $\cC$ the allone and the allzero codewords we
obtain the Hadamard code~\cite[p. 49]{MWS}.
\end{example}

\begin{example}
\label{ex:ooc} Let $\C$ be the $[n, 2k,k]_q$
code of size $\frac{q^n-1}{q^k-1}$ defined above.
By applying Construction FDTW on $\C$ we
obtain a $(q^n, 2 \cdot q^k -2 ,q^k)$
code $\cC$ of size $q^{n-k} \frac{q^n-1}{q^k-1}$
which is a Steiner system $S(2,q^k,q^n)$.
\end{example}

For the analysis of the next two  families of
optimal codes (see Theorem~\ref{thm:new_codes})
we need the following two theorems.
The first one is the well-known Johnson bound~\cite{Jon62}.
The second theorem was developed in~\cite{AVZ}.

\begin{theorem}
\label{thm:Jh1} If $n \geq w > 0$ then $$A(n,d,w) \leq
\floorenv{\frac{n}{w} A(n-1,d,w-1) }$$.
\end{theorem}
\vspace{-0.2cm}
\begin{theorem}
\label{thm:AVZ} If $b > 0$ then
$$
A(n,2 \delta , w) \leq \floorenv{ \frac{\delta}{b}}~,
$$
where
$$
b= \delta - \frac{w(n-w)}{n} + \frac{n}{M^2} \left\{ M \frac{w}{n}
\right\} \left\{ M \frac{n-w}{n} \right\}
$$
$$
M=A(n,2\delta,w),~~~
\{ x \} = x - \floorenv{x}~.
$$
\end{theorem}

\vspace{0.1cm}

The next theorem presents two new optimal constant weight codes
derived by shortening codes obtained via Construction FDTW.
To obtain large constant weight codes via
construction FDTW large constant dimension codes are required
and hence constructions of large constant dimension codes
are required. One such construction which produces codes
used in the next theorem is the multilevel construction
introduced in~\cite{EtSi09}. The code used in the construction
is derived also from equation~(\ref{eq:part_spread}) which follows.
\begin{lemma}
\label{lem:for_new_codes}
$\cA_q [2m-1,2m-2,m] = q^m +1~.$
\end{lemma}

\begin{theorem}
\label{thm:new_codes}
$~$
\begin{itemize}
\item $A(2^{2m-1}-1, 2^{m+1}-4, 2^m -1) = 2^m +1$.

\item $A(2^{2m-1}, 2^{m+1}-4, 2^m) = 2^{2m-1} +2^{m-1}$.
\end{itemize}
\end{theorem}
\begin{proof}
The upper bound $A(2^{2m-1}-1, 2^{m+1}-4, 2^m -1) \leq 2^m +1$ is
a direct application of theorem~\ref{thm:AVZ}. Using this bound in
Theorem~\ref{thm:Jh1} we obtain the second upper bound
$A(2^{2m-1}, 2^{m+1}-4, 2^m) \leq 2^{2m-1} +2^{m-1}$.

By applying Construction FDTW on a $[2m-1,2^m+1 ,2m-2,m]_2$
code~(see Lemma~\ref{lem:for_new_codes}) we obtain a $(2^{2m-1},
2^{m+1}-4, 2^m)$ code of size~$2^{2m-1} +2^{m-1}$.
Hence, $A(2^{2m-1}, 2^{m+1}-4, 2^m) \geq
2^{2m-1} +2^{m-1}$ and thus $A(2^{2m-1}, 2^{m+1}-4, 2^m) =
2^{2m-1} +2^{m-1}$. By shortening the $(2^{2m-1}, 2^{m+1}-4, 2^m)$
code of size $2^{2m-1}+2^{m-1}$ we obtain a $(2^{2m-1}-1,
2^{m+1}-4, 2^m-1)$ code of size $2^m +1$ and hence $A(2^{2m-1}-1, 2^{m+1}-4, 2^m
-1) = 2^m +1$.
\end{proof}

Construction FDTW requires large constant dimension codes. But,
usually even the largest constant dimension codes will not induce
large constant weight codes via Construction FDTW. The examples we
have given in this section represent three classes of constant
dimension codes from which large constant weight codes will be
formed via Construction FDTW, where by large we mean, close enough
to the value of $A(n,d,w)$. These three classes are:

\begin{enumerate}
\item $[n,2k,k]_q$ codes.

\item $[n,n-1,n-1]_2$ codes.

\item $[n,2k-2,k]_2$ codes.
\end{enumerate}

For the first class of constant dimension codes,
it was proved in~\cite{EtVa11} that if $n \equiv r\! \pmod k$.
then, for all $q$, we have
\begin{equation}
\label{eq:part_spread}
\cA_q(n,2k,k) \:\ge\: \frac{q^n\! -\,
q^k(q^r\!-1)-1}{q^k-1}~.
\end{equation}
By applying construction FDTW on the
related code we obtain a $(q^n ,2 \cdot q^k -2 ,q^k)$ code
of size $\frac{q^{2n-k}\! -\,
q^n(q^r\!-1)-q^{n-k}}{q^k-1}$, while the
related upper bound is $A(q^n , 2 \cdot q^k -2 , q^k ) \leq
\floorenv{q^{n-k} \floorenv{\frac{q^n -1}{q^k -1}}}$.
There are some known minor improvements to this upper bound.
The second class is small in
its size. For the third class, we can use codes obtained by the
various known constructions. But, we believe that
larger constant dimension codes of this class can be found.
Some constant weight codes
obtained from these codes by Construction FDTW
can be of size not far from the related
upper bounds.

We will consider now optical orthogonal codes.
An $(n,w,\lambda)$ {\it optical orthogonal code} $\cC$ is a set of
codewords (each codeword is a $w$-subset) with the following
properties:

\begin{itemize}
\item Each codeword has length $n$ and weight $w$.

\item If $X \in \cC$ then all the $n$ cyclic shift of $X$. $X$ is their only
representative in $\cC$.

\item If $X'$ and $Y'$ be any cyclic shifts of $X, Y \in \cC$, $X'
\neq Y'$, then $|X' \cap Y'| \leq \lambda$, where $X$ and $Y$
are taken as $w$-subsets.
\end{itemize}

Optical orthogonal codes were considered in many papers,
e.g.~\cite{CSW,AGM,OMK,MOKL}. We
will now show how to use construction FDTW to form optical
orthogonal codes. For this purpose, we will define the concept of
cyclic code in $\cG_q (n,k)$. Let $\alpha$ be a primitive element
of $\GF(q^n)$. We say that~a~code $\C \subseteq \cG_q(n,k)$ is
{\it cyclic} if it has the following property: whenever
$\{\zero,\alpha^{i_1},\alpha^{i_2},\ldots,\alpha^{i_m}\}$ is a
codeword of $\C$, so is its cyclic~shift $\{\zero,
\alpha^{i_1+1},\alpha^{i_2+1},\ldots,\alpha^{i_m+1} \}$. In other
words, if we map each vector space $V \,{\in}\, \C$ into the
corresponding binary characteristic vector of length $q^n-1$
(excluding the {\it zero} element) then the set of all such
characteristic vectors is closed under cyclic shifts. Note that
the property of being cyclic does \emph{not} depend on the choice
of a primitive element $\al$ in $\GF(q^n)$. The proof of the following
lemma is simple and from lack of space it is left for the reader.

\begin{lemma}
If $\C$ is a cyclic code then the codes $\cC_0^{q^n-1}$ and
$\cC_1^{q^n-1}$ are cyclic, where $\cC$ is the code obtained
from $\C$ by Construction FDTW.
\end{lemma}
%\begin{proof}
%%$\cC_1^{q^n-1}$ is a cyclic code by the definition of a cyclic
%%code $\C$, the definition of shortened code, and the definition of
%%a cyclic codes over $\F_q$.
%Let $\alpha$ be a primitive element in $\GF(q^n)$ and assume that
%$\{ 0 , \alpha^{i_1} , \alpha^{i_2} , \ldots , \alpha^{i_{q^n-1}}
%\} \in \C$. If $0 \leq j \leq q^n-2$, then $ch(\{ 0 , \alpha^{i_1}
%, \alpha^{i_2} , \ldots , \alpha^{i_{q^n-1}} \}) \in \cC$ and
%$ch(\{ \alpha^j , \alpha^j + \alpha^{i_1} , \alpha^j +\alpha^{i_2}
%, \ldots , \alpha^j +\alpha^{i_{q^n-1}} \}) \in \cC$. Since $\C$
%is a cyclic code it follows that $ch(\{ 0 , \alpha^{i_1+1} ,
%\alpha^{i_2+1} , \ldots , \alpha^{i_{q^n-1}+1} \} \in \cC$ and
%$ch(\{ \alpha^{j+1} , \alpha^{j+1} + \alpha^{i_1+1} , \alpha^{j+1}
%+ \alpha^{i_2+1} , \ldots , \alpha^{j+1} + \alpha^{i_{q^n-1}+1} \}
%\in \cC$. Therefore, $\cC_0^{q^n-1}$ and $\cC_1^{q^n-1}$ are
%cyclic.
%\end{proof}

\vspace{0.1cm}

Kohnert and Kurz~\cite{KoKu08}, Etzion and Vardy~\cite{EtVa11} have
considered $[n,4,3]_2$ cyclic codes. Some of the
codes have the following parameters: An $[8,4,3]_2$ code of size 1275 (compared
to $\cA_2 (8,4,3) \leq 1493$); $[9,4,3]_2$ code of size 5694
($\cA_2 (9,4,3) \leq 6205$); $[10,4,3]_2$ code of size 21483
($\cA_2 (10,4,3) \leq 24698$). The first two codes are the largest possible
cyclic code with their parameters. The resulting constant weight
codes obtained by Construction FDTW have the following parameters: $(256
,12,8)$ code of size 40800 (compared to $A(256,12,8) \leq 48960$); $(512
, 12,8)$ code of size 364416  (compared to $A(512,12,8) \leq 397120$);
$(1024 , 12,8)$ code of size 2749824 (compared to $A(1024,12,8) \leq
3180032$). Given an $(n,d,w)$ cyclic constant weight code $\cC$ we form an optical
orthogonal code as follows. We partition the codewords into
equivalence classes such that two codewords are in the same
equivalence class if one can be formed from the other by a cyclic
shift. From each equivalence class of size $n$ we take one
representative to form the optical orthogonal code.
For the above cyclic codes have the
following parameters of optical orthogonal codes:
$(255,7,1)$ and size 1275; $(255,8,2)$ and
size 38525; $(511,7,1)$ and size 5621; $(511,8,2)$ and size
354123; $(1023,7,1)$ and size 21483; $(1023,8,2)$ and size
2728341. Similarly, optical orthogonal codes are obtained by
shortening the codes of example~\ref{ex:ooc}. These codes
coincide with the codes in~\cite{OMK,MOKL}.

%@@@@@@@@@@@@@@@@@@@@@@@@@@@@@@@@@@@@@@@@@@@@@@@@@@@@@@@@@@@@@@@@@@@@@@@@%
%                                                                        %
%   5. Nonexistence of Nontrivial Perfect Codes                          %
%                                                                        %
%@@@@@@@@@@@@@@@@@@@@@@@@@@@@@@@@@@@@@@@@@@@@@@@@@@@@@@@@@@@@@@@@@@@@@@@@%
\section{Encoding, Decoding, and Error-Correction}
%\vspace{-.25ex}
\label{sec:encode}

Unfortunately, most known large constant
weight codes do not have efficient encoding and decoding
algorithms. The same is true for an efficient error-correction
algorithm. It appears that if the constant weight code is
constructed via Construction FDTW from a constant dimension code
which has efficient encoding and decoding algorithms then
efficient encoding and decoding algorithms can be designed also for
the constant weight code obtained via
Construction FDTW. In the sequel we need
the reduced row echelon form of a subspace.

\subsection{Reduced row echelon form}
%\vspace{-.25ex}
Let $X\in \Gr$ be a $k$-dimensional subspace.
We can represent $X$ by the $k$ linearly independent vectors from
$X$ which form  a unique $k\times n$ generator matrix in {\it
reduced row echelon form} (RREF), denoted by $RE(X)$, and defined
as follows:
\begin{itemize}
\item The leading coefficient of a row is always to the right of
the leading coefficient of the previous row.

\item All leading coefficients are {\it ones}.

\item Every leading coefficient is the only nonzero entry in its
column.
\end{itemize}

For each $X\in \Gr$  we associate a binary vector of length $n$
and weight $k$, $v(X)$,
where the \textit{ones} in $v(X)$ are exactly in the
positions where $RE(X)$ has the leading \textit{ones}.

Let $\cI (X)$ be the set of $n-k$ positions numbers in $v(X)$ with
{\it zeroes}. Let $CP(X)$ be an $(n-k) \times n$ binary matrix
with rows of weight one. The set of positions of the {\it ones} in
these rows is exactly $\cI (X)$. Note, that the $k$ rows of
$RE(X)$ together with the $n-k$ rows of $CP(X)$ span $\F_q^n$.

\subsection{Encoding and decoding}
%\vspace{-.25ex}
Let $\C$ be an $[n,d=2t,k]_q$ code with an efficient encoding
algorithm EA. Construction FDTW yields a
$(q^n , 2 \cdot q^k - 2 \cdot q^{k-t} ,q^k)$
code $\cC$ of size $q^{n-k} | \C |$. We can consider the set $\{
(i,j) ~:~ i \in \Z_M ,~ j \in \F_q^{n-k} \}$, where $M = | \C |$, as the set of
information words for the code $\cC$ (since $M$ is the number of
codewords in $\C$ and from each codeword of $\C$ we derive~$q^{n-k}$
codewords in $\cC$). The encoding algorithm for an
information word $(i,j)$ is straightforward. First, we encode $i$
to a $k$-dimensional subspace $X =\{ 0 , \alpha_1 , \ldots ,
\alpha_{q^k-1} \}$ by the algorithm EA. Let $B(j)$ be the row
vector of length $n-k$ which forms the $q$-ary representation of
$j$. We encode the information word $(i,j)$ to the binary codeword
$ch( B(j) \cdot CP(X) + X)$ which has weight $q^k$. Note, that
$B(j) \cdot CP(X)$ is the vector used to form the appropriate
coset of $X$. It is not necessarily a coset leader, but
it represents the coset in the encoding.

Decoding of a codeword into an information word is done similarly
in reverse order. What we need for this algorithm is a constant
dimension code with an efficient encoding algorithm. For this
purpose we can use the constant dimension codes generated by
lifting of rank-metric codes~\cite{SKK,EtSi09}.

\subsection{Error-correction}
%\vspace{-.25ex}
In this subsection we will consider the codewords of the constant weight
codes as the elements of the finite field from which the characteristic
vector was constructed. We note that when an $(n,d,w)$ code is
used, both codewords and the received words are vectors of length $n$ and
weight $w$.

Again, let $\C$ be an $[n,d=2t,k]_q$ code from which Construction
FDTW yields a $(q^n ,2 \cdot q^k - 2 \cdot q^{k-t} ,q^k)$ code
$\cC$ of size $, q^{n-k} |\C|$.
As we should assume that the received words also have
weight $q^k$, the code $\cC$ is capable to correct if
at most $q^k - q^{k-t} -2$ errors occurred (at most
$\frac{q^k - q^{k-t} -2}{2}$ {\it ones} were changed to {\it zeroes}, and
vice versa, in a codeword which can be recovered).
However, we will concentrate only on the error-corrections
capabilities of the codes due to the fact that the codewords
are characteristic vectors of $k$-dimensional subspaces or their cosets.
Hence, we will assume that less than~$\frac{q^k}{2}$ errors occurred.

For simplicity we will consider the codewords as $q^k$-subsets of
$\F_q^n$, i.e., the elements of $\F_q^n$ from which the codewords
of $\C$ were formed.
Assume that the codeword $X=\{x_1, x_2 , \ldots ,x_{q^k}\}$
was transmitted and the word $Y=\{y_1, y_2 , \ldots ,y_{q^k}\}$ was
received. We start by generating the multiset $\cT (Y)$ of the
2-subsets differences from $Y$, i.e., $\cT (Y) = \{ y_i - y_j ~:~ 1 \leq
i < j \leq q^k \}$. Note, that if $Y$ is a codeword or a coset then
all these subtractions results in elements of the codeword since
a codeword is a linear subspace. Note also that if $q$ is even then the order
of the two elements is a substraction does not change the result.
This implies the distinction in the sequel between $q$ even and $q$ odd.
$|\cT (Y)| =\binom{q^k}{2}$ if $q$ is even
and $|\cT (Y)| =q^{2k} - q^k$ if $q$ is odd. Let $z_1 , z_2
, \ldots , z_{q^k}$ be the elements with the most appearances in
$\cT$. We form the codeword $Z=\{ z_1, z_2 , \ldots ,z_{q^k} \}
\in \C$. Let $\beta \in Y$ be any element that was used at least
$\frac{3 \cdot q^k}{4}$ times to form elements from $Z$,
i.e., $z_{i_r} = y_{i_r} - \beta$, where $z_{i_r} \in Z$ and
$y_{i_r} \in Y$. If less than $\frac{q^k}{2}$ errors
occurred then the submitted codeword is $ch(\beta + Z) = ch(\{ \beta
+z_1, \beta +z_2 , \ldots ,\beta +z_{q^k} \})$.

the correctness of this error-correction algorithm is based on the
following two lemmas.

\begin{lemma}
Let $\C$ be an $[n,d,k]_q$ constant dimension code. Let $\cC$ be a
$(q^n , 2 \cdot q^k - 2 \cdot q^{k-t} ,q^k)$ code generated by
Construction FDTW and let $X=\{ \alpha_1 , \alpha_2 , \ldots ,
\alpha_{q^k-1} \} \in \cC$. Then
\begin{enumerate}
\item An element which appears in $\cT (X)$ has $\frac{q^k}{2}$
appearances in $\cT (X)$ if $q$ is even and $q^k$ appearances
if $q$ is odd.

\item Assume that due to errors, $\tau$ {\it zeroes} were changed
to {\it ones} and $\tau$ {\it ones} were changed to {\it zeroes}
in $X$, and a word $Y$ was formed. Then an element which appears
in $\cT (X)$ has at least $\frac{q^k}{2} - \tau$ appearances in $\cT
(Y)$ if $q$ is even and $q^k - 2\tau$ appearances in $\cT (Y)$ if $q$ is odd.

\item Assume that due to errors, $\tau$ {\it zeroes} were changed
to {\it ones} and $\tau$ {\it ones} were changed to {\it zeroes}
in $X$, and a word $Y$ was formed. Then an element which does not
appear in $\cT (X)$ has at most $\tau$ appearances in $\cT (Y)$
if $q$ is even and at most $2 \tau$ appearances in $\cT (Y)$
if $q$ is odd.

\item For each $\beta \in \F_q^n$ we have $\cT (X) = \cT (\beta +X)$.
\end{enumerate}
\end{lemma}

\begin{lemma}
Let $\C$ be an $[n,d,k]_q$ constant dimension code. Let $\cC$ be a
$(q^n  ,2 \cdot q^k - 2 \cdot q^{k-t} ,q^k)$ code generated by
Construction FDTW and $X=\{ \alpha_1 , \alpha_2 , \ldots ,
\alpha_{q^k} \} \in \cC$ formed from the codeword $Z=\{ \gamma_1 ,
\gamma_2 , \ldots , \gamma_{q^k} \} \in \C$, i.e., $X = \beta + Z$
for some $\beta \in \F_q^n$. Then
\begin{enumerate}
\item Each element of $X$ is used to form each one of the elements
of $Z$ in $\cT(X)$ (the elements of $Z$ and $\cT(X)$
coincide, and each element of $Z$ appears exactly $\frac{q^k}{2}$ times
in $\cT(X)$ if $q$ is even and $q^k$ times if $q$ is
odd).

\item Assume that there were $2 \tau$
errors and a word $Y$ was formed. Each element of $Y$ which
appears also in $X$ is used to form at least $\frac{q^k}{2} - \tau$ elements
of $Z$ in $\cT(Y)$ if $q$ is even and at least $q^k - 2\tau$ elements
of $Z$ in $\cT(Y)$ if $q$ is odd.

\item Assume that there were $2 \tau$
errors and a word $Y$ was formed. Each element of $Y$ which does
not appear in $X$ is used to form at most $\tau$ elements of $Z$
in $\cT(Y)$ if $q$ is even and at most $2\tau$ elements of $Z$
in $\cT(Y)$ if $q$ is odd.
\end{enumerate}
\end{lemma}
\begin{cor}
The error-correction algorithm can recover any codeword obtained
from Construction FDTW if less than $\frac{q^k}{2}$ errors occurred.
\end{cor}

\vspace{-0.05in}
%@@@@@@@@@@@@@@@@@@@@@@@@@@@@@@@@@@@@@@@@@@@@@@@@@@@@@@@@@@@@@@@@@@@@@@@@%
%                                                                        %
%   6. Conclusion and Open Problems                                      %
%                                                                        %
%@@@@@@@@@@@@@@@@@@@@@@@@@@@@@@@@@@@@@@@@@@@@@@@@@@@@@@@@@@@@@@@@@@@@@@@@%
\section{Conclusion}
\vspace{-.25ex} \label{sec:last}

We have presented a construction for a constant weight code from a
given constant dimension code. Some of the constructed codes are either
optimal or the largest known constant weight codes.
The main advantage of the new codes is that they have
efficient algorithm for error-correction; and if there exists an
efficient encoding/decoding algorithms for the related constant
dimension code then also the constant weight code has efficient
encoding/decoding algorithms.

The error-correction algorithm used only the fact that all codewords
were constructed from distinct subspaces. One direction of research is
to design an efficient error-correction algorithm which will be able to correct
all errors if no more than $q^k - q^{k-t} -2$ errors occurred.
More connections between constant
weight codes and constant dimension codes should be
also be explored.

\vspace{-0.2cm}

\section*{Acknowledgment}
The work  was supported in part by the U.S.-Israel
Binational Science Foundation, Jerusalem, Israel,
Grant No. 2012016.

\vspace{-0.0cm}

%\bibliography{allbib,extra}

\end{document}